\documentclass[12pt]{article}
\newcommand{\beq}{\begin{equation}}
\newcommand{\eeq}{\end{equation}}

\newcommand{\beqn}{\begin{eqnarray}}
\newcommand{\eeqn}{\end{eqnarray}}

\newcommand{\Ga}{\mbox{${\Gamma}$}}

\newcommand{\br}{\mbox{${\mathbf r}$}}

\newcommand{\al}{\mbox{${\alpha}$}}
\newcommand{\la}{\mbox{${\lambda}$}}

\begin{document}
\begin{center}
{\bf Hydrogen atom in strong magnetic field revisited\\}
\vspace*{1cm} I.B. Khriplovich and G.Yu. Ruban
\\ {\it Budker Institute of Nuclear Physics,\\} {\it 630090
Novosibirsk, Russia,\\} {\it and Novosibirsk University }
\end{center}

\vspace*{.5cm}
\begin{abstract}
We derive in a straightforward way the spectrum of a hydrogen atom
in a strong magnetic field.
\end{abstract}

\vspace*{1cm} The spectrum of hydrogen atom in a strong magnetic
field was found long ago~\cite{el}, and is presented now in
textbooks (see, e.g.,~\cite{ll} (\S 112, Problem 3)). The approach
used therein is as follows. At first, using the wave functions of
electron in a magnetic field, one constructs an effective
potential for the motion along the field, and then the spectrum in
this potential is found. We present here a somewhat different
solution of the problem, with the results coinciding in fact with
those of~\cite{el,ll}. We hope however that our approach,
physically straightforward and transparent, is of some interest.
At least, the present note may be considered as a sort of
mini-review on the subject.

Our starting point is the obvious observation that in a
sufficiently strong magnetic field $H$ (the exact criteria are
discussed below) the motion of an atomic electron becomes almost
one-dimensional, along the magnetic field, in the Coulomb
potential $- e^2/z$.

The corresponding one-dimensional wave equation is
\beq\label{we}
u^{\prime\prime} +
\left(-\,\frac{1}{4}\,+\,\frac{\nu}{z}\right)u=0.
\eeq
We have introduced in it the usual dimensionless variable:
\[
\frac{2z}{a\nu}  \rightarrow z;
\]
here $a=\hbar^2/m_e e^2$ is the Bohr radius, $m_e$ is the electron
mass, $\nu$ is the effective quantum number, related to the
electron energy as
\beq
E_{\nu}= - \,\frac{m_e e^4}{2\hbar^2 \nu^2}\,.
\eeq
Equation (\ref{we}) coincides exactly with the radial equation for
the $s$-wave in the three-dimensional Coulomb potential $-e^2/r$,
and has therefore the common hydrogen spectrum
\beq\label{sp-}
E_n^- =-\,\frac{m_e e^4}{2\hbar^2 n^2}\, \quad n=1,2,3 ... \,,
\eeq
and the set of solutions
\beq\label{so-}
u_n^-(z)=\exp{(-z/2)}\, z \,F(-n,2;z)\,, \quad \nu = n\,,
\eeq
where $F$ is the confluent hypergeometric function (here and below
we are not interested in the normalization factors). These
solutions vanish at the origin and are trivially continued to $z <
0$. Thus obtained solutions on the whole $z$ axis are odd under $z
\rightarrow -z$ (as reflected by the superscripts ``minus'' in
(\ref{sp-}), (\ref{so-})).

There is however an essential difference between the present
problem and the $s$-wave Coulomb one. In the last case (\ref{so-})
is the only solution. The reason is well-known. Naively the radial
wave equation for $R(r)(= u(r)/r)$ has two independent solutions,
which behave for $r \rightarrow 0$ as $R \sim$ const ($u \sim r$)
and $R \sim 1/r$ ($u \sim$ const), respectively. However, in fact
$R \sim 1/r$ is no solution at all for the homogeneous wave
equation if the point $r=0$ is included, since $\triangle (1/r) =
- 4\pi \delta (\br)$. As to our problem, equation (\ref{we}) does
not describe really the vicinity of $z=0$ since therein we have to
consider seriously the magnetic field itself. Therefore, there are
no reasons to discard those solutions of (\ref{we}) which tend to
a constant for small $z$ (and of course decrease exponentially for
$z \rightarrow \infty$).

Such solutions are presented in a convenient form in \cite{ll}
(Mathematical Appendices, \S~d, (d.17)). To our purpose they can
be written for $z> 0$ as
\[
u_{\nu}^+(z)=\exp{(-z/2)}\,\Biggl\{ 1 - \nu z  \Biggl(\ln z
\,F(1-\nu,2;z)\Biggr.\Biggr.
\]
\beq
\label{so+} \Biggl. \Biggl.
+\sum_{k=0}^{\infty}\, \frac{\Ga(1-\nu + k)\,[\psi (1-\nu +k) -
\psi(k+2) - \psi(k+1)]}{\Ga(1-\nu)\, k! \,(k+1)!}\, z^k \Biggr)  \Biggr\};
\eeq
here $\psi(\al)$ denotes the logarithmic derivative of the gamma
function: $\psi(\al)= \Ga^{\prime}(\al)/\Ga(\al)$. Being trivially
continued to $z < 0$, thus obtained solutions on the whole $z$
axis are even under $z \rightarrow -z$ (as reflected by the
superscript ``plus'' in (\ref{so+}) and in the corresponding
eigenvalues below).

Under any reasonable regularization of the logarithmic singularity
at $z \rightarrow 0$, the even solutions should have vanishing
first derivative at the origin. In this way we obtain the
following equation for the eigenvalues of $\nu$:
\beq\label{eq}
\ln \frac{a}{a_H}\, = \,\frac{1}{2\nu}\, + \psi (1-\nu);
\eeq
here $a_H = \sqrt{\hbar c/eH}$ is the typical scale for the radius
of electron orbits in the magnetic field $H$. We are working in
the logarithmic approximation, i.e. assume that
\beq\label{la}
\la = \ln a/a_H \gg 1.
\eeq
This allows us to use a crude cut-off at $a_H$ for the formal
logarithmic divergence at $z \rightarrow 0$, as well as to
simplify somewhat this equation.

The smallest root of equation (\ref{eq}) is
\beq
\nu_0^+ =\,\frac{1}{2\la}\,,
\eeq
which gives the ground state energy
\beq\label{0}
E_0^+ =-\,\frac{m_e e^4}{2\hbar^2}\, \ln^{-2}\left(\frac{\hbar^3
H^2}{m_e^2 e^3 c}\right) \,.
\eeq
Other roots of equation (\ref{eq}) are
\beq
\nu_n^+ = n + \,\frac{1}{\la}\, \quad n=1,2,3 ... \,,
\eeq
with the corresponding energies
\beq\label{sp+}
E_n^- =-\,\frac{m_e e^4}{2\hbar^2
n^2}\,\left[1-\frac{2}{n}\,\ln^{-1}\left(\frac{\hbar^3 H^2}{m_e^2
e^3 c}\right)\right] \,.
\eeq

Let us mention that the one-dimensional Coulomb problem was
considered in [3 -- 5] with various regularizations of the
singularity at $z \rightarrow 0$, but without any relation to the
problem of the hydrogen atom in a strong magnetic field.

But let us come back to our problem. The resulting spectrum of the
hydrogen atom in a strong magnetic field looks as follows. Each
Landau level in this field serves as an upper limit to the
sequence of discrete levels of the Coulomb problem in the $z$
direction. This discrete spectrum consists of a singlet ground
state with the energy given by formula (\ref{0}), and close
doublets of odd and even states of energies given by formulae
(\ref{sp-}) and (\ref{sp+}). There is also a continuous spectrum
of the motion along $z$ above each Landau level.

This picture is valid for sufficiently low Landau levels, as long
as the radius of a magnetic orbit is much less than the Bohr
radius. Obviously, in a strong magnetic field this description
fails for large magnetic quantum numbers, i.e., in the
semiclassical region. Here we can estimate the orbit radius
directly from the well-known spectrum of electron in a magnetic
field (see, e.g.,~\cite{ll} (\S 112, Problem 1)):
\beq\label{spl}
E = \hbar \,\frac{eH}{m_e c}\,(N+1/2), \quad
N=n_{\rho}+\frac{m+|m|}{2}\,;
\eeq
here $n_{\rho}$ is the radial quantum number in the $xy$ plane,
and $m$ is the angular momentum projection onto the $z$ axis. Now
the semiclassical estimate for the magnetic radius is
\[
a_H(N)\approx \sqrt{\frac{\hbar c}{eH}\,N}= a_H\sqrt{N}\,.
\]
Thus the present picture of levels holds as long as
\beq
\la = \ln\frac{a}{a_H} \gg \ln N.
\eeq

At last, let us consider the correspondence between the obtained
system of levels in a strong magnetic field and the hydrogen
spectrum in a vanishing field. A beautiful solution of this
problem was given in \cite{kl} (and is quoted in \cite{el}). We
would like to present the solution here as well (having in mind in
particular that our note can be considered as a mini-review).

The crucial observation made in \cite{kl} is as follows. While
changing the magnetic field from vanishingly small to a very
strong one, the number of nodal surfaces of a given wave function
remains the same. A hydrogen wave function (in zero magnetic
field) with quantum numbers $n$, $l$, $m$ has $n_r=n-l-1$ nodal
spheres and $l-|m|$ nodal cones with a $z$ axis. With the increase
of the magnetic field, the nodal spheres become ellipsoids of
rotation, more and more prolate, tending to cylinders in the limit
of infinite field. The correspondence between $n_r$ and $n_{\rho}$
($n_{\rho}$ being the radial quantum number in the $xy$ plane in a
strong magnetic field) gets obvious from this picture:
\beq\label{rro}
n_{\rho}=n_r.
\eeq
The evolution of the hydrogen nodal cones is less obvious.
However, due both to equation (\ref{rro}) and to the conservation
of the total number of nodal surfaces, the number $n_z$ of the
nodes of a solution of equation (\ref{we}) should coincide with
the number of nodes of the corresponding spherical function,
\beq\label{nlm}
n_z=l-|m|.
\eeq
In other words, $l-|m|$ nodal cones of a hydrogen wave function
evolve into $n_z$ planes of constant $z$ corresponding to the
nodes of an eigenfunction of equation (\ref{rro}).

And at last, let us note that during the whole evolution of the
magnetic field, $m$ remains constant.

Let us consider now, for instance, the ground state in the
magnetic field, with $N=0$ (see (\ref{spl})). Obviously, it is
degenerate, and its corresponding magnetic wave functions have
$n_{\rho}=0$ and $m=\,0,\,-1,\,-2\; ...\,$. Let us confine further
to its lowest sublevel corresponding to the ground state solution
of equation (\ref{we}), with $n_z=0$. According to the above
arguments, the hydrogen ancestors of those wave functions should
have $n_r =0$ and $l=|m|$. In other words, these ancestors are:
\[
1s; \quad 2p,\,m=-1; \quad 3d,\,m=-2; \quad {\rm and\;\; so\;\;
on}.
\]

\vspace*{1.0 cm}

We are grateful to V.A. Novikov, V.V. Sokolov, M.I. Vysotsky, and
A.V. Zolotaryuk for useful discussions. We acknowledge the support
by the Russian Foundation for Basic Research through grant No.
03-02-17612.

\end{document}